\documentclass{PoS}

\usepackage{natbib}
\usepackage[english]{babel}
\usepackage{mycommands}
\usepackage{graphicx}
\usepackage{amssymb,amsmath}
\usepackage{sidecap}

\usepackage{wrapfig}
\bibpunct{(}{)}{;}{a}{}{,}     % A&A citation style
\title{Star Formation: Chemistry as a Probe of Embedded Protostars}

\ShortTitle{Star Formation: Chemistry as a Probe of Embedded Protostars}

\author{\speaker{Ruud Visser}\\
        Department of Astronomy, University of Michigan, 500 Church Street, Ann Arbor, MI 48109-1042, USA\\
        E-mail: \email{visserr@umich.edu}}

\abstract{The embedded phase of star formation is the crucial phase where most of the stellar mass is assembled. Velocity-resolved spectra reveal an infalling envelope, bipolar outflows, and perhaps an infant circumstellar disk -- all locked together in a cosmic dance of gravitational collapse and magnetic winds. Densities and temperatures change by orders of magnitude as the protostar evolves, driving a chemistry as exotic as it is fascinating. I will review two examples of how to exploit chemistry and molecular spectroscopy to study the physics of low-mass star formation: energetic feedback and episodic accretion.}

\FullConference{Frank N. Bash Symposium 2013: New Horizons in Astronomy (BASH 2013)\\
		October 6-8, 2013\\
		Austin, Texas }

\begin{document}

\section{Introduction}
\label{sec:intro}
\noindent
For all of human history, we have wondered about the origins of the Earth, the Sun, and the rest of the solar system. The notion that stars form from the gravitational collapse of a molecular cloud core traces back to the ``nebular hypothesis'' developed by \citet{swedenborg1734a}, \citet{kant1755a}, and \citet{laplace1796a}. Since then, many generations of astronomers have filled in the details of what such cloud cores look like, how they collapse, and how this results in a planetary system \citep{cameron88a}. Others have studied the differences between low-mass and high-mass stars, between stars born in isolation or in clusters, and between star formation in the Milky Way or other galaxies \citep{kennicutt12a}.

The state of affairs in the field of star formation has been reviewed periodically, in part through the \emph{Protostars and Planets} series of conferences. The most recent edition, PPVI, took place in July 2013 in Heidelberg, Germany and featured chapters on molecular clouds \citep{dobbs14a,andre14a}, embedded protostars \citep{li14a,dunham14a}, chemical evolution \citep{vandishoeck14a,ceccarelli14a}, and much more. Earlier reviews on various aspects of star formation were written by, e.g., \citet{shu87a}, \citet{vandishoeck93a}, \citet{vandishoeck98a}, \citet{lada99a}, \citet{andre00a}, \citet{klein07a}, \citet{white07a}, and \citet{mckee07a}.

\begin{figure}[t!]
\centering
\includegraphics[width=\textwidth]{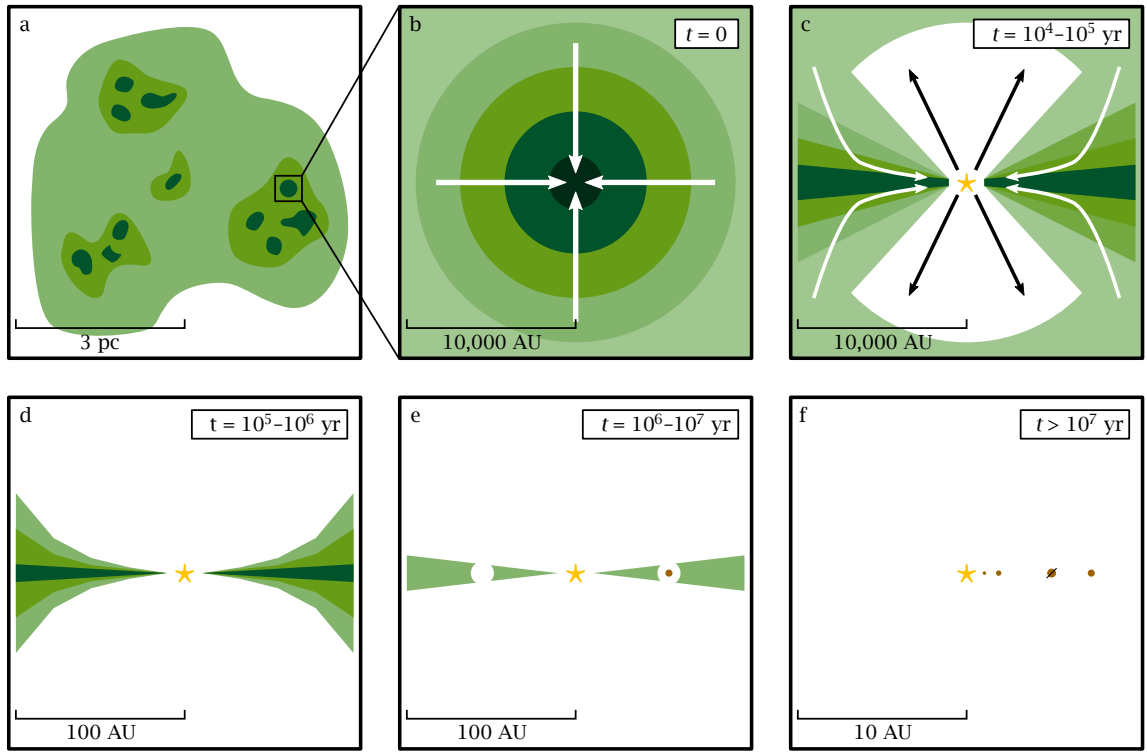}
\caption{Schematic view of isolated low-mass star formation, based on \citet{shu87a}. Darker colors represent higher densities. The typical sizes and timescales for each step are indicated.}
\label{fig:sfpanels}
\end{figure}

The process of low-mass star formation is usually split into four to six steps, as illustrated in \fig{sfpanels}. The first step (panel a) is a molecular cloud of typically a few pc in size with a mean density of about 100 \pcc{} and a total mass of a few 1000 \msun{} \citep{bergin07b}. The cloud contains denser clumps of a few 100 \msun{} each, which in turn contain individual dense cores (\ten{4}--\ten{5} \pcc) of a few \msun{} in mass and about 0.1 pc in size. In many star-forming regions, the clumps and cores line up in filaments \citep{andre14a}.

Star formation properly starts when a dense core loses pressure support and begins to undergo gravitational collapse (\fig{sfpanels}b). Due to the core's rotation, material falling in from the outer parts gets deflected towards the gravitational midplane. The young star soon launches a bipolar jet that carries away some of the core's angular momentum. The jet carves out a bipolar cavity from the infalling envelope and drives an outflow into the surrounding molecular cloud \citep[\fig{yso-parts};][]{bally07a,ray07a,arce07a,shang07a,frank14a}. Initially, when the young star is still deeply embedded in its natal core, it is called a Class 0 or Stage 0 protostar (\citealt{andre93a}; see \citealt{robitaille06a} for the distinction between observational classes and evolutionary stages). The transition to Stage I (\fig{sfpanels}c) occurs when the star has accreted enough matter that its mass exceeds that of the remnant circumstellar reservoir \citep{lada99a,robitaille06a}. The median lifetimes for Stage 0 and I protostars -- together called the embedded stage -- are 0.10 and 0.44 Myr \citep{evans09a}.

After the envelope is cleared away, we are left with a pre--main-sequence T Tauri star surrounded by a circumstellar disk of gas and dust \citep[\fig{sfpanels}d;][]{bertout89a,dullemond07b}. Although disks of one form or another have now been detected in Stage 0 and I sources \citep[e.g.,][]{jorgensen07a,takakuwa12a,tobin12a,murillo13a}, it remains unclear how and when disks first form and how big they grow in the embedded phase \citep{li14a}. In T Tauri stars, disks can be anywhere from less than 10 AU to a few 100 AU in radius and their mass can be as high as 0.1 \msun{} \citep{andrews07b}. They last for a few Myr, during which the dust settles towards the midplane and grows first to larger grains, then to rocks and boulders, and ultimately to planets \citep{johansen14a,raymond14a,helled14a,benz14a}. Meanwhile, the gas in the disk gradually gets dispersed into the interstellar medium \citep{alexander14a} or accreted by giant planets. The combined effects of planet formation and gas dispersal can leave gaps and holes, producing a so-called transitional disk \citep[\fig{sfpanels}e;][]{espaillat14a}. After the gas is cleared away, the system turns into a debris disk in which rocky planets can continue to grow \citep{matthews14a}. This phase can last for another 10-100 Myr before we end up with a main-sequence star and a mature planetary system (\fig{sfpanels}f).

The current review covers two topics of recent interest in the embedded phase of star formation: energetic feedback (\sect{feedback}) and episodic accretion (\sect{epacc}). Both parts discuss in particular the use of chemistry and molecular spectroscopy as a probe of the the underlying physical processes.

% ------------------------------------------------------------

\section{Energetic feedback}
\label{sec:feedback}
\noindent
Star formation is a violent process. As the cloud core collapses and material accretes onto the young star, the gravitational potential energy of the infalling gas is released back into the protostellar environment \citep{mckee07a}. Nuclear fusion within the star forms an additional source of energy output; for low-mass stars, this begins at about 0.1 Myr after the onset of collapse \citep{stahler83a,myers98a}.

Feedback is important in regulating the overall star-formation efficiency, i.e., the fraction of total cloud mass that actually makes it into stars \citep{shu87a,shu04a,evans09a}. On galactic scales, feedback from massive stars shapes the interstellar medium through ionizing radiation and stellar winds \citep{chu08a,arthur08a,dopita08a,oey08a}. Low-mass protostars exert much the same type of feedback on their environment, but on much smaller scales: they affect the physics and chemistry of the circumstellar disk and envelope through processes like photoelectric heating, ionization, and shock compression \citep{visser12a}. In numerical simulations of low-mass star formation, the higher temperatures due to radiative feedback enhance the stability of the disk (reducing fragmentation) and produce a better fit to the initial mass function \citep{offner10a,bate12a}.

The warm and hot gas (from 100 to a few 1000 K) in low-mass protostars emits mainly at mid- to far-infrared wavelengths. Every escaping photon carries away a quantum of energy; the dominant ``cooling lines'' are pure rotational transitions of \mh, CO, \w, and OH \citep{giannini01a,vandishoeck04a,neufeld09a,karska13a}. Over the past decade, the Spitzer Space Telescope \citep[5--37 \micron;][]{werner04a} and the Herschel Space Observatory \citep[55--670 \micron;][]{pilbratt10a} have provided a wealth of data, building on earlier pioneering work with the Infrared Space Observatory \citep[ISO; 2.5--200 \micron;][]{kessler96a}. For example, Spitzer covered the lowest eight rotational lines of \mh{} with upper-level energies up to 7200 K \citep{neufeld09a}, along with many lines of \w{} (700--3000 K) and some of OH \citep[400--3000 K;][]{watson07a}. Herschel filled in the lower-excitation range of \w{} and OH, and covered the CO rotational ladder up to the $J=49$--48 transition with $\eup=6700$ K \citep{herczeg12a}.

\begin{figure}[t!]
\centering
\includegraphics[width=0.707\textwidth]{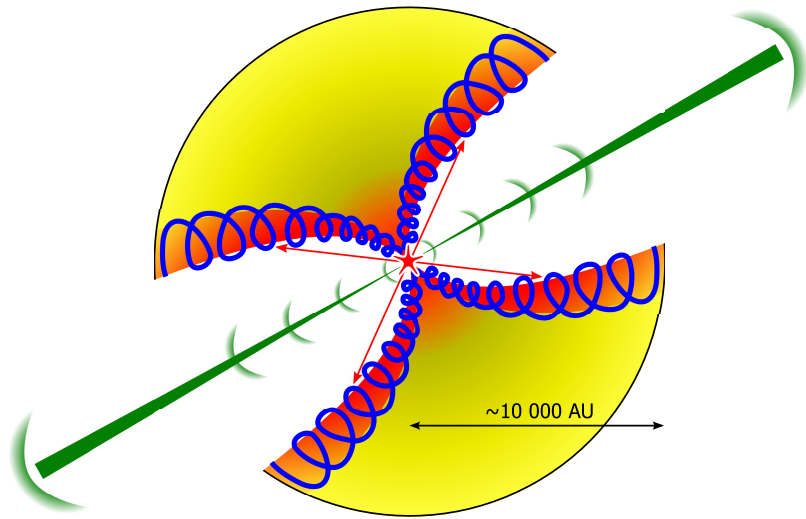}
\caption{Schematic representation of a low-mass embedded protostar from \citet{visser12a}. The gas temperature in the infalling envelope ranges from 10 K at the outer edge (yellow) to a few 100 K close to the star (red). The protostellar UV field illuminates the gas in the walls of the bipolar outflow cavity and heats it to a few 100 K even at large distances (red). Shocks driven by the jet (green) or the stellar wind can heat the gas in the cavity walls to a few 1000 K.}
\label{fig:yso-parts}
\end{figure}

The cooling lines observed with Spitzer and Herschel can originate in various parts of the protostar (\fig{yso-parts}). Quantifying and disentangling the contributions from each physical component has been a major challenge, not in the least because most of the observations are spatially and spectrally unresolved.\footnote{\ Towards the short end of their respective wavelength ranges, Spitzer and Herschel had a resolution of 500--2000 AU in nearby star-forming regions. The notable exception to the spectrally unresolved observations are the data from the heterodyne spectrometer HIFI on Herschel \citep{degraauw10a}.} I will discuss the CO rotational lines in more detail as an example of both the opportunities and the limitations from the Herschel observations.

Rotationally excited CO is ubiquitous in low-mass embedded protostars. Early observations with ground-based sub-millimeter telescopes and with ISO measured line intensities from $J=6$--5 up to 21--20 that far exceed any reasonable predictions for just an infalling envelope heated by the protostellar luminosity \citep{spaans95a,hogerheijde98a,giannini01a,nisini02a}. Two alternative heating mechanisms were suggested: ultraviolet photons and shocks.

The PACS spectrometer on Herschel had access to nearly all CO lines from $J=13$--12 up to 49--48 \citep{poglitsch10a}. The three key programs DIGIT, HOPS, and WISH combined to observe the CO ladder in more than 50 low-mass embedded protostars \citep{vankempen10b,vankempen10a,visser12a,herczeg12a,goicoechea12a,manoj13a,karska13a,green13a,dionatos13a,lee13a,lindberg14a}. A common way to plot these data is in a rotational diagram, with the upper-level energies for the observed lines on the horizontal axis and the population of each rotational level on the vertical axis \citep{goldsmith99a}. For optically thin emission and gas in local thermodynamic equilibrium (LTE), the points in such a diagram fall on a straight line. The slope of this line is then the inverse of the gas temperature.

\begin{figure}[t!]
\centering
\includegraphics[width=0.6\textwidth]{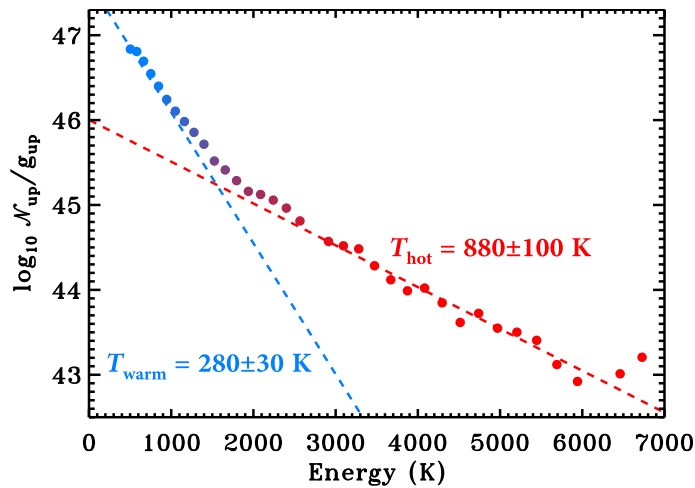}
\caption{Rotational diagram for CO in the Class 0 protostar NGC1333 IRAS4B, adapted from \citet{herczeg12a}. The vertical axis plots the total number of detected molecules for each $J$ level divided by the degeneracy of that level.}
\label{fig:iras4b-co}
\end{figure}

For sources with a sufficient number of detected lines, the rotational points generally do \emph{not} fall on a straight line and instead require two temperature components for a good fit \citep[\fig{iras4b-co};][]{herczeg12a,manoj13a,karska13a,green13a}. The breakpoint falls at $\eup\approx1800$ K in all sources, between the 25--24 and 26--25 transitions. Although the full source sample spans more than two orders of magnitude in bolometric luminosity, the two temperature components are remarkably constant: $320\pm50$ K (``warm'') and $850\pm150$ K (``hot'').\footnote{\ The missing ``cold'' component dominates the CO lines up to about $J=10$--9 \citep{yildiz13a}.} What do these temperatures mean and why do they show so little variation across more than 50 protostars?

One possible explanation is that the two temperature fits reflect two different reservoirs of CO gas. Detailed radiative transfer models constructed for NGC1333 IRAS2A, HH46, and DK Cha were able to reproduced the CO intensities with a combination of UV- and shock-heated gas \citep{vankempen10a,visser12a}. Based in part on spectrally resolved CO 16--15 data from Herschel-HIFI, \citet{kristensen13a} concluded that the ``hot'' component originates in dense gas ($>10^6$ \pcc) within 100 AU of the central star. The UV-heated gas appears to be more spread out, up to a few 1000 AU, but still has to have similarly high densities to maintain LTE\@. Therein lies the weakness of this scenario: why would a sample of more than 50 protostars, spanning more than two orders of magnitude in $\lbol$, always have UV-heated gas of about 300 K?

An alternative explanation is that the CO emission originates in subthermally excited gas (non-LTE) with a density of less than \ten{6} \pcc{} \citep{neufeld12a,manoj13a}. The gas can either be uniformly hot ($>2000$ K) or feature a continuous range of temperatures from a few 10 to a few 1000 K\@. Such conditions are consistent with shocks along the outflow or the cavity walls at several 100 to 1000 AU from the central star. The advantage of the non-LTE solution is that it does not require a constant gas temperature across the source sample; instead, a fairly wide range of temperatures can all produce the same rotational excitation conditions. However, it is unclear whether the column density of hot, low-density gas is sufficient to match the observed line intensities. Invoking shocked gas as the sole origin of the high-$J$ CO emission is also inconsistent with the narrow line profiles seen in velocity-resolved spectra of CO 6--5, 10--9, and 16--15, indicative of a reservoir of quiescent hot gas (\citealt{vankempen09b}; \citealt{yildiz12a,yildiz13a}; Kristensen et al. in prep.).

% ------------------------------------------------------------

\section{Episodic accretion}
\label{sec:epacc}

% --------------------------------------------------

\subsection{Solving the luminosity problem}
\label{sec:lprob}
\noindent
In the simplest theories of star formation, there is a constant flow of matter from the collapsing envelope onto the star and the stellar mass grows at a constant rate $\dot{M} \equiv \el{d}M_\ast/\el{d}t$ \citep{shu77a}. The stellar luminosity scales with the mass accretion rate and with the ratio of the stellar mass over the stellar radius \citep{adams85a}:
\begin{equation}
\label{eq:tau}
L_\ast \approx \frac{GM_\ast\dot{M}}{R_\ast}\,.
\end{equation}
The constant flow from these simple theories continues until the infalling envelope is depleted at the end of the embedded phase. T Tauri stars show some ongoing accretion from the disk onto the star, but at a rate that is two or three orders of magnitude slower than in Class 0 and I sources.

\begin{figure}[t!]
\centering
\includegraphics[width=0.707\textwidth]{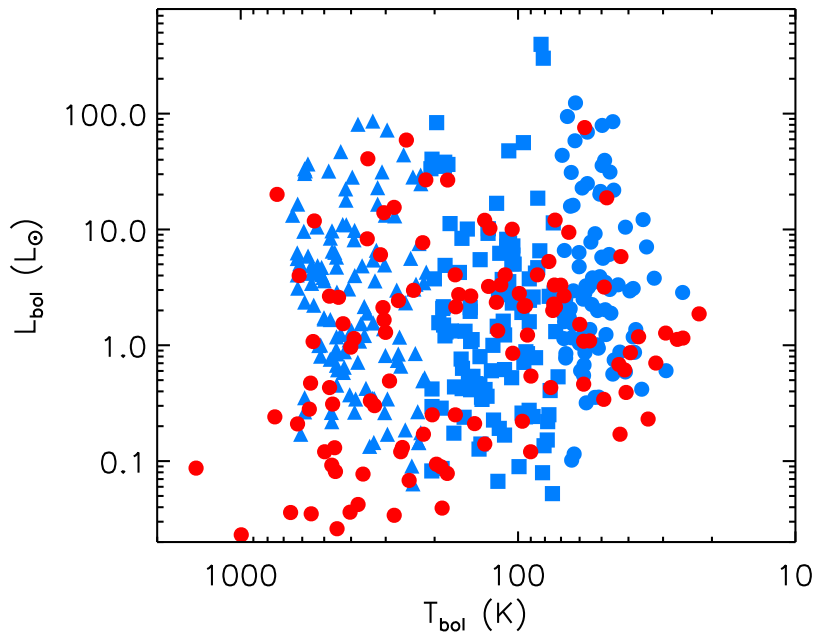}
\caption{Bolometric temperatures and luminosities for the embedded protostars from \citet[red]{evans09a} and \citet[blue]{fischer13a}. The former are mixed Class 0 and I; the latter are split into Class 0 (circles), early Class I (squares), and late Class I (triangles).}
\label{fig:blt}
\end{figure}

If the accretion rate is indeed constant throughout the embedded phase, a sample of low-mass protostars at the same evolutionary state should have roughly the same luminosity. Furthermore, embedded protostars should be more luminous than T Tauri stars. However, surveys with the Infrared Astronomical Satellite (IRAS), Spitzer, and Herschel have all shown that neither prediction holds true \citep{kenyon90a,kenyon94a,kenyon95a,dunham08a,enoch09b,evans09a,fischer13a}. For example, \fig{blt} plots the bolometric luminosities versus the bolometric temperatures\footnote{\ $\tbol$ is the temperature of a blackbody with the same mean frequency as the full spectral energy distribution (SED) from optical to millimeter wavelengths \citep{myers93a}. It generally increases as a protostar evolves \citep{chen95a}. $\lbol$ is the integrated flux across the SED; for embedded protostars, it is essentially equal to $L_\ast$.} for several hundred Class 0 and I sources observed with Spitzer and Herschel. The luminosities span more than three orders of magnitude at any given bolometric temperature (i.e., evolutionary state), which cannot be explained by a constant mass accretion rate onto the star.

Two possible solutions have been proposed for this ``luminosity problem'': an accretion rate that decreases with time, or a low accretion rate for most of the time interspersed by short bursts of high accretion \citep{kenyon90a}. Accretion bursts are known to occur in T Tauri stars \citep[FUor and EXor events;][]{herbig77a,hartmann96a,audard14a}. Similar bursts in embedded sources should be detectable at mid-IR, far-IR, and sub-mm wavelengths \citep{johnstone13a}. A survey of 4000 protostars with a five-year baseline identified 1--4 burst candidates \citep{scholz13a}. The presence of periodic shocks along protostellar jets also suggests that burst of some sort probably occur \citep{reipurth89a,raga02a}. Radiative transfer calculations find better agreement with the $\lbol$ and $\tbol$ observations from \fig{blt} for scenarios with accretion bursts than they do for a steadily declining accretion rate \citep{myers98a,offner11a,dunham12a}. The ``episodic accretion'' scenario finds further support in numerical simulations, where mass accretes from the envelope onto the disk until the disk becomes unstable and unloads onto the star in a short burst of high accretion \citep{vorobyov05b,vorobyov06a,vorobyov09a,vorobyov10b,zhu09a,zhu10b}. Based on kinematic observations, the Class 0 protostar NGC1333 IRAS4A appears to be in a state of mass loading \citep{mottram13a}.

% --------------------------------------------------

\subsection{Chemical tracers of episodic accretion}
\label{sec:epchem}
\noindent
Because of the difficulty of directly observing embedded accretion bursts, some groups are looking into the use of chemistry and molecular spectroscopy as an alternative tracer. Each accretion burst causes an increase in luminosity, which heats up the entire circumstellar envelope. The dust temperature responds particularly fast to a change in luminosity: $<100$ s at 10 AU away from the star up to \scit{5}{6} s at a typical outer radius of \ten{4} AU \citep{johnstone13a}. The gas heats up more slowly and reaches a new equilibrium temperature after $\sim$\ten{5} s at 10 AU and $\sim$\ten{9} s or $\sim$30 yr at \ten{4} AU\@. That last value is comparable to the expected duration of $\sim$100 yr for the strongest accretion bursts \citep{vorobyov05b,dunham12a}.

The relevant chemistry is driven by freeze-out and evaporation. When a neutral gas-phase molecules hits a dust grain, it sticks to the surface if the temperature $T$ is below some critical value called the evaporation temperature. This value varies from one molecule to the next; in a protostellar envelope, $\tevap\approx20$ K for CO and 100 K for \w{} \citep{fraser01a,bisschop06a}. The molecules sticking to the grain surface form an ice mantle. If a cold dust grain is heated up, the ices sublimate back into the gas according to their respective evaporation temperatures.

When the envelope heats up during an accretion burst, the evaporation timescale is less than a year wherever the new temperature exceeds $\tevap$ for some molecule \citep{hasegawa92a,visser12b}. When the burst ends and the envelope cools back down, the freeze-out timescale is much longer: \ten{4} yr at a typical gas density of \ten{6} \pcc. This results in a prolonged period during which the gas and dust temperatures in the envelope are characteristic of a quiescent state of accretion, while the chemistry is still representative of the hot phase from the most recent burst. In other words, the quiescent phase features abundance patterns inconsistent with its low luminosity.

\begin{wrapfigure}{r}{0.42\textwidth}
\centering
\vspace{-0.2cm}
\includegraphics[width=0.4\textwidth]{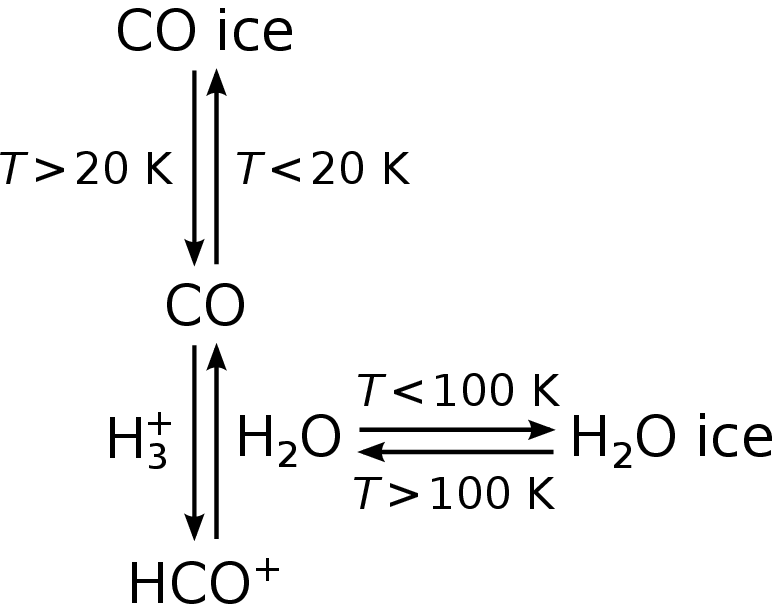}
\vspace{-0.3cm}
\end{wrapfigure}

Commonly observed species like CO, \hcop, and \nnhp{} are potential tracers of episodic accretion, because their abundances are sensitive to changes in temperature \citep{lee07a,visser12b,vorobyov13a}. For example, the chemistry of \hcop{} can be boiled down to the set of three reversible reactions shown on the right. The dominant formation channel is the reaction between CO and \hhhp, which gets suppressed below 20 K due to freeze-out of CO\@. Above 100 K, \hcop{} is destroyed mainly by \w\@. At lower temperatures, \w{} is frozen out and electrons become the main destruction channel (not shown).

\begin{figure}[t!]
\centering
\includegraphics[width=\textwidth]{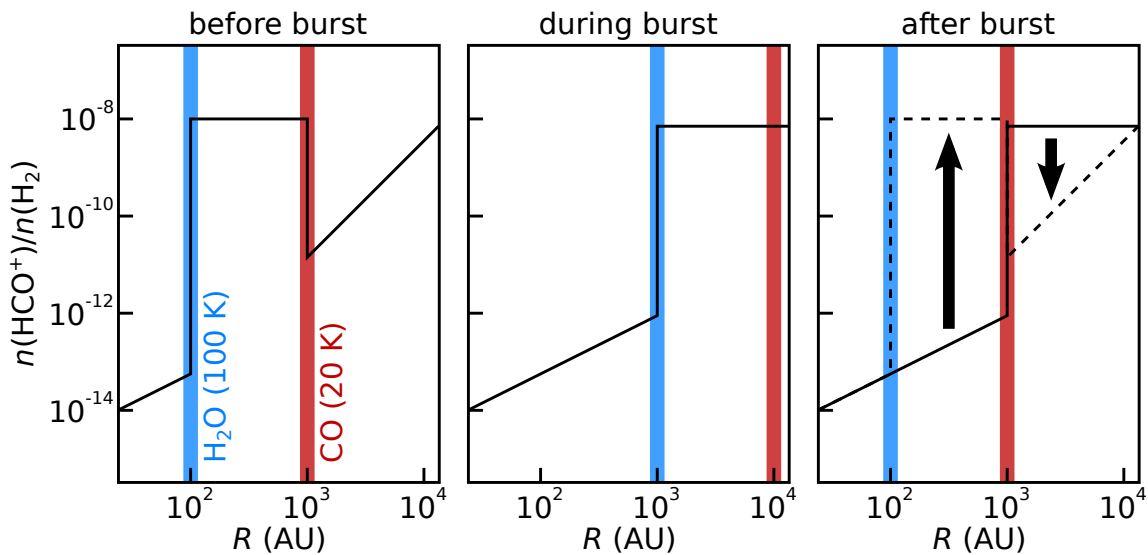}
\caption{Schematic radial abundance profiles for \hcop{} in a protostellar envelope before, during, and after an accretion burst. The radii corresponding to the evaporation temperatures of \w{} ice (100 K) and CO ice (20 K) are marked with blue and red bars. Based on Visser \& Bergin (in prep.).}
\label{fig:hcop-epacc}
\end{figure}

\figg{hcop-epacc} shows the schematic radial abundance profile for \hcop{} before, during, and after a burst, based on the chemical models of Visser \& Bergin (in prep.). Before the burst, the profile breaks down into three components: a low inner abundance due to destruction by \w{} above 100 K; a high abundance between 20 and 100 K, where CO is in the gas phase and \w{} is not; and a lower outer abundance because CO is depleted onto the grains.\footnote{\ Due to the density gradient in the envelope, the freeze-out timescale of CO increases towards the outer edge. Hence, the CO and \hcop{} abundance both increase with radius in the part of the envelope below 20 K.} The higher luminosity during the burst heats up the envelope and moves the \w{} and CO evaporation radii outwards. The envelope quickly cools down after the burst \citep{johnstone13a}, but the chemistry can only reset itself according to the much longer timescales associated with freeze-out at different radii \citep{visser12b}. Until that happens, there is an excess of cold \hcop{} ($<20$ K) and a lack of warm \hcop{} (20--100 K) relative to the current dust temperature profile.

One way to observe such an anomalous \hcop{} abundance profile would be through single-dish spectra of a low- and a high-excitation line, such as $J=3$--2 and 7--6 (Visser \& Bergin in prep.). Spatially resolved interferometric observations offer even better constraints. For example, \citet{jorgensen13a} targeted the $J=4$--3 line of the optically thin isotopolog \htcop{} in the Class 0 protostar IRAS 15398 with the Atacama Large Millimeter Array (ALMA). The emission shows up as a ring around the central star. The dust temperature inferred at the inner edge of the ring is only 30 K, too low for \w{} to be abundant in the gas and destroy \hcop. Hence, it appears the temperature was higher in the past and the chemistry is still in the process of readjusting to the current cold environment. The temperature discrepancy suggests a burst that raised the stellar luminosity by a factor of 100. Based on the freeze-out timescale of \w, this burst happened 100--1000 yr ago \citep{jorgensen13a}.

Other chemical evidence for episodic accretion comes in the form of Spitzer observations of the \cdo{} ice band at 15 \micron. The shape and width of this spectral feature depend on whether the \cdo{} ice is pure or mixed with CO and/or \w{} \citep{ehrenfreund97a,vanbroekhuizen06a}. In protostellar envelopes, \cdo{} ice forms out of CO ice below 20 K\@. Pure \cdo{} ice -- known by a characteristic double peak in the 15-\micron{} band -- is produced by heating the grains to 20--40 K to evaporate CO\@. \cdo{} is more tightly bound and remains frozen. This purification step is irreversible \citep{hagen83a}, so if pure \cdo{} ice is detected below 20 K, it is an indication of a warmer past \citep{lee07a}. In embedded protostars with luminosities of less than 1 \lsun, the bulk of the envelope is too cold to form pure \cdo{} ice at detectable levels. Nonetheless, \citet{kim11a,kim12a} detected a clear double-peaked line profile in the \cdo{} ice band for six out of 19 such low-luminosity sources, and concluded that they must have experienced an accretion burst at some point. Unlike the \htcop{} data from \citet{jorgensen13a}, the \cdo{} ice bands offer no information about when these bursts occurred.

% ------------------------------------------------------------

\section{Conclusions}
\label{sec:conc}
\noindent
Despite decades of work, many aspects of low-mass star formation remain poorly understood. Recent observations with Spitzer and Herschel have shed new light on the energy balance in embedded protostars and on the relative importance of feedback from UV photons and shocks. Another topic of current interest is episodic accretion, i.e., a scenario in which a young star gains most of its mass in short bursts while spending most of the time in a state of low accretion. Chemistry and molecular spectroscopy are proving to be invaluable tools in the worldwide efforts to characterize and understand energetic feedback and episodic accretion in more detail.

%\begin{thebibliography}{99}
%\bibitem{...} 
%....
%
%\end{thebibliography}

\bibliographystyle{apj}
\bibliography{bash}

\end{document}